\documentclass{aa}
\usepackage{graphicx,natbib}
\newcommand{\HETE}{\mbox{HETE-2}}
\newcommand{\Swift}{\textit{Swift}}
\newcommand{\Chandra}{\textit{Chandra}}

\usepackage{txfonts}
\begin{document}
\title{Optical emission from GRB\,050709: a short/hard GRB in a star-forming galaxy%
\thanks{Based on observations carried out at ESO telescopes under
programmes Id 075.D-0787 and 075.D-0468.}}

\author{S. Covino\inst{1} \and D. Malesani\inst{2} \and G.L. Israel\inst{3} 
\and P. D'Avanzo\inst{1,4} \and L.A. Antonelli\inst{3} \and G. Chincarini\inst{1,5} 
\and D. Fugazza\inst{1} \and M.L. Conciatore\inst{3} \and M. Della Valle\inst{6} 
\and F. Fiore\inst{3} \and D. Guetta\inst{3} \and K. Hurley\inst{7} \and D. Lazzati\inst{8} 
\and L. Stella\inst{3} \and G. Tagliaferri\inst{1} \and M. Vietri\inst{9} \and S. Campana\inst{1} 
\and D.N. Burrows\inst{10} \and V. D'Elia\inst{3} \and P. Filliatre\inst{11,12} \and 
N. Gehrels\inst{13} \and P. Goldoni\inst{11,12} \and A. Melandri\inst{3,14} \and 
S. Mereghetti\inst{15} \and I.F. Mirabel\inst{16} \and A. Moretti\inst{1} \and 
J.A. Nousek\inst{10} \and P.T. O'Brien\inst{17} \and L.J. Pellizza\inst{12} \and 
R. Perna\inst{8} \and S. Piranomonte\inst{3} \and P. Romano\inst{1} \and F.M. Zerbi\inst{1}}

\offprints{S. Covino, \email{covino@mi.astro.it}}

\institute{ 
INAF, Osservatorio Astronomico di Brera, via E. Bianchi 46, I-23807 Merate (Lc), Italy
\and        
International School for Advanced Studies (SISSA-ISAS), via Beirut 2-4, I-34014 Trieste, Italy
\and        
INAF, Osservatorio Astronomico di Roma, via di Frascati 33, I-00040 Monteporzio Catone (Roma), Italy
\and        
Dipartimento di Fisica e Matematica, Universit\`a dell'Insubria, via Valleggio 11, I-22100 Como, Italy
\and        
Universit\`a degli studi di Milano-Bicocca, Dipartimento di Fisica, piazza delle Scienze 3, I-20126 Milano, Italy
\and        
INAF, Osservatorio Astrofisico di Arcetri, largo E. Fermi 5, I-50125 Firenze, Italy
\and        
University of California, Berkeley, Space Sciences Laboratory, Berkeley, CA 94720-7450, USA
\and       
JILA, University of Colorado, 440 UCB, Boulder CO 80309-0440, USA
\and      
Scuola Normale Superiore, piazza dei Cavalieri 7, I-56126 Pisa, Italy
\and       
Department of Astronomy \& Astrophysics, Pennsylvania State University, State College, PA 16801, USA
\and      
Laboratoire Astroparticule et Cosmologie, UMR 7164, 11 Place Marcelin Berthelot, F-75231 Paris Cedex 05, France 
\and      
Service d'Astrophysique, DSM/DAPNA, CEA Saclay, F-91911 Gif-sur-Yvette Cedex, France
\and      
NASA, Goddard Space Flight Center, Greenbelt, MD 20771, USA
\and      
Universit\`a degli Studi di Cagliari, Dipartimento di Fisica, I-09042 Monserrato (Ca), Italy
\and      
INAF/IASF Milano ``G. Occhialini'', via E. Bassini 15, I-20133 Milano, Italy
\and      
European Southern Observatory - Vitacura, Casilla 19001, Santiago 19, Chile 
\and      
X-Ray \& Observational Astronomy Group, Dept. of Physics \& Astronomy, University of Leicester, Leicester LE1 7RH, UK
}
\date{}

\abstract{

We present optical observations of the short/hard gamma-ray burst
GRB\,050709, the first such event with an identified optical
counterpart. The object is coincident with a weak X-ray source and is
located inside a galaxy at redshift $z = 0.1606 \pm 0.0002$. Multiband
photometry allowed us to study the broad-band spectral energy
distribution.
Late-time monitoring places strong limits on any supernova simultaneous
with the GRB. The host galaxy is not of early type. Spectra show that
the dominant stellar population is relatively young ($\sim 1$~Gyr), and
that ongoing star formation is present at a level of
$2\mbox{--}3\,L/L_*~M_\odot$/yr. This is at least 2 orders of magnitude
larger than that observed in the elliptical hosts of the short
GRB\,050509B and GRB\,050724. This shows that at least some short GRBs
originate in a young population. Short/hard GRB models based on the
merger of a binary degenerate system are compatible with the host galaxy
characteristics, although there is still the possibility of a connection
between young stars and at least a fraction of such events.
}

\authorrunning{Covino et al.}
\titlerunning{GRB\,050709 in a star forming galaxy}

\maketitle
%

\section{Introduction}
\label{sec:intro}

\begin{figure*}
\includegraphics[width=\textwidth]{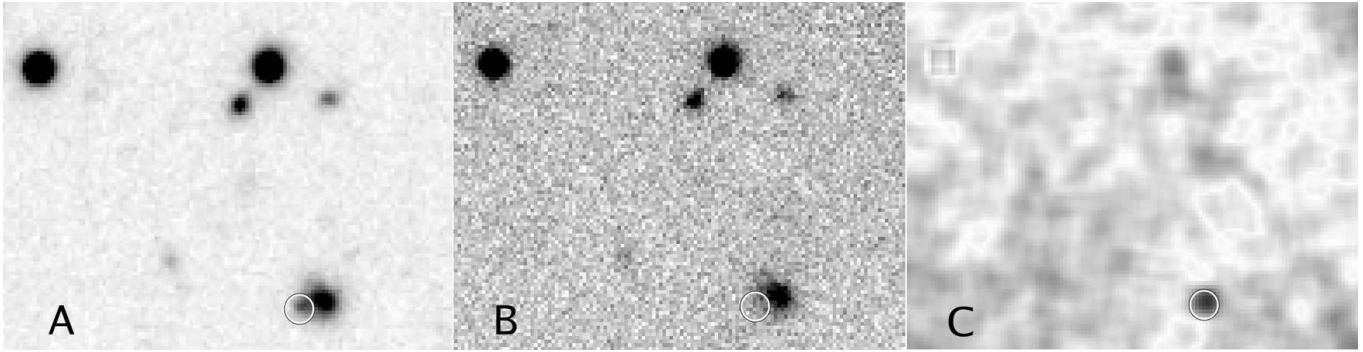}
\caption{$R$-band image of the field of GRB\,050709, on 2005 Jul 12.41
(A) and 20.42 (B). Panel C shows the result of the subtraction,
evidencing a fading source coincident with the \Chandra{} counterpart
(circle). The boxes cover a region $30\arcsec \times 20\arcsec$ wide.%
\label{fig:gal}}
\end{figure*}

Gamma-ray bursts (GRBs) are short pulses of gamma rays occurring at
random positions in the sky. Two classes of GRBs are currently known,
characterised by different durations and spectral properties
\citep{Kou93}. Long GRBs (typically lasting 10--100~s) are on the
average softer than the short ones (duration $< 2$~s). Over the past
years, great advances have been made in understanding the former class,
thanks to the observation of their optical and radio
counterparts. However, until recently, no optical emission from short
GRBs had been identified \citep{Hur02}, leaving fundamental questions
about their nature, progenitors and distances unanswered.

Recently, thanks to the \Swift{} and \HETE{} satellites, accurate and
rapid localisations of short GRBs have become available, enabling deep,
sensitive searches at long wavelengths. \Swift{} discovered a weak X-ray
counterpart to GRB\,050509B \citep{Geh05}, located $\sim 11\arcsec$ away
from a bright elliptical galaxy at $z = 0.2248$
\citep{Blo05,Hjo05,CT05}. An early-type galaxy \citep{Goro05} is also
associated with GRB\,050724 \citep{Ba05} at $z = 0.257$ \citep{Pro05},
for which also an optical afterglow was singled out inside the host
\citep{Beral05,Dava05}. Finally, the line of sight of GRB\,050813
\citep{Ret05} lies towards a galaxy cluster at $z = 0.722$.  A few
elliptical galaxies were identified inside the XRT error circle, again
supporting the association with early-type galaxies
\citep{Gla05,Ber05,Pro05}.

GRB\,050709 was discovered by \HETE{} on 2005 Jul 9.94209 UT
\citep{Vill05}. Its prompt emission consisted of a single pulse lasting
70~ms in the 3--400 keV band, followed by a weaker, soft bump $\sim
100$~s long. This second episode may be due to the afterglow onset
\citep{Vill05}, or to flaring activity \citep{Ba05,King05,Perna05}. In
any case, the prompt emission properties are consistent with those
of a short/hard GRB.

Follow-up observations with the \Chandra{} X-ray observatory revealed a
faint, uncatalogued X-ray source inside the \HETE{} error circle
\citep{Fox05a}. At the coordinates $\alpha_{\rm J2000} = 23^{\rm h}
01^{\rm m} 26\fs9$, $\delta_{\rm J2000} = -38\degr 58\arcmin
39\farcs5$ (0\farcs4 error), it was coincident with a pointlike object
embedded in a bright galaxy \citep{Jen05} at $z = 0.16$
\citep{Pri05a}. The variability of this source led \citet{Pri05b} to
propose it as the optical counterpart of GRB\,050709.

\section{Observations and data analysis}\label{sec:data}

\begin{table}
\caption{
Observation log and photometry of the transient source. Errors are at the
1$\sigma$ confidence level, while upper limits are at 3$\sigma$. Data
were not corrected for Galactic extinction.%
\label{tab:mag}}
\centering
\begin{tabular}{ccrcr}
\hline
\textbf{Date}    & \textbf{Instrument} &\textbf{Exp.} &\textbf{Filter} &\textbf{Magnitude}\\
(UT)             &                     &(min)         &                &                  \\
\hline
2005/07/12 09:44 & FORS\,2             & 6            & $V$            & $24.38 \pm 0.10$ \\
2005/07/14 07:21 & FORS\,1             & 6            & $V$            & $> 25.00$        \\
2005/07/20 10:16 & FORS\,1             & 6            & $V$            & (reference)      \\
2005/07/30 02:37 & FORS\,2             & 9            & $V$            & $> 25.20$        \\
2005/07/12 09:57 & FORS\,2             & 5            & $R$            & $23.83 \pm 0.07$ \\
2005/07/20 10:07 & FORS\,1             & 6            & $R$            & (reference)      \\
2005/07/30 02:54 & FORS\,2             & 50           & $R$            & $> 25.00$        \\
2005/07/12 09:32 & FORS\,2             & 10           & $I$            & $> 23.25$        \\
2005/07/14 07:32 & FORS\,1             & 5            & $I$            & $> 24.10$        \\
2005/07/18 06:38 & FORS\,1             & 20           & $I$            & (reference)      \\
2005/07/30 04:10 & FORS\,2             & 9            & $I$            & $> 23.50$        \\
\hline
\end{tabular}
\end{table} 

We observed the field of GRB\,050709 with the ESO Very Large Telescope
(VLT), using the FORS\,1 and FORS\,2 instruments. In our first images,
taken on 2005 Jul 12, the pointlike object reported by \citet{Jen05} was
clearly visible in the $V$ and $R$ bands (Fig.~\ref{fig:gal}). Its
coordinates are $\alpha_{\rm J2000} = 23^{\rm h} 01^{\rm m} 26\fs96$,
$\delta_{\rm J2000} = -38\degr 58\arcmin 39\farcs3$ (0\farcs25 error),
fully consistent with the \Chandra{} position. The object lies inside a
bright galaxy ($\approx 1\farcs2$ away from its nucleus), whose
magnitudes are $V = 21.35 \pm 0.07$, $R = 21.08 \pm 0.07$, and $I =
20.63 \pm 0.08$. To search for brightness variations, we monitored the
field at several epochs. Data analysis was performed by adopting a
subtraction technique \citep{AlLu98}, well suited to identify variable
objects even when blended with nearby sources. The pointlike object was
found to be variable (at the $\sim 10\sigma$ level), being undetectable
from Jul 14 onwards. This confirms the independent finding of
\citet{Hjorth05b}. Magnitudes of the variable source were computed
assuming a negligible flux in the reference epoch (see
Table~\ref{tab:mag}). Photometry of the transient was performed by
inserting artificial stars of known brightness and calibrated by
observing Landolt standard fields.

On Jul 30 we took medium-resolution spectra of the host
galaxy. Observations were carried out with the FORS\,2 instrument at the
VLT-UT1, with the 300V grism, covering the wavelength range
6000--9200~\AA{} (6~\AA{} FWHM). From the detection of several emission
lines, among them H$\alpha$, H$\beta$, and [O\,II], we derived a
redshift $z = 0.1606 \pm 0.0002$. This is consistent with the results of
\citet{Fox05b}. Therefore, the rest-frame $B$-band luminosity of the
host is%
\footnote{Assuming a cosmology with $\Omega_{\rm m} = 0.3$,
$\Omega_\Lambda = 0.7$, and $h_0 = 0.71$.}
$L_{\rm B} \sim 3.5 \times 10^{42}$~erg~s$^{-1}$ \citep[$\sim 0.10 L_*$,
assuming $M_B^* = -20.13$ as determined from the SDSS
survey;][]{Blanton03} and the candidate afterglow lies at a projected
distance of $\approx 3.3$~kpc from the galaxy core.

\section{Discussion}\label{sec:disc}

\begin{figure}
\includegraphics[width=\columnwidth]{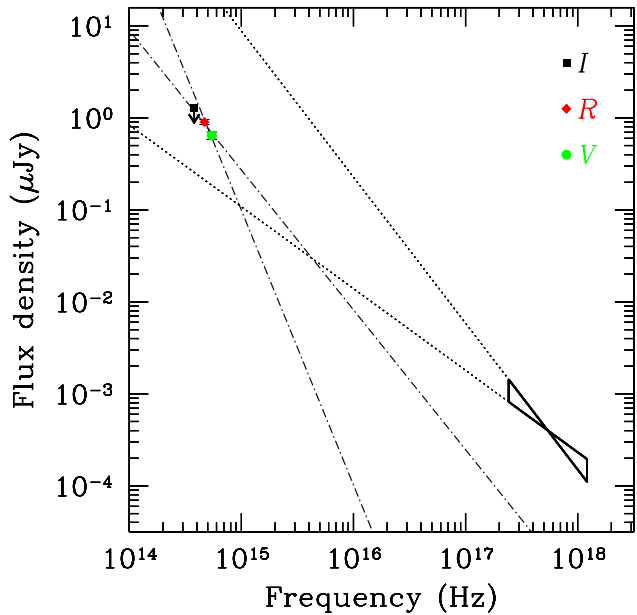} 
\caption{Broad-band spectral energy distribution of the afterglow of
GRB\,050709 on Jul 12.4 UT. The bow-tie shaped region represent the
simultaneous X-ray spectrum taken from \Chandra{} \citep{Fox05b}. The
dot-dashed and dotted lines indicate the extrapolation of the optical
and X-ray spectra, respectively.%
\label{fig:sed}}
\end{figure}

Assuming a power-law flux decay ($F(t) \propto t^{-\alpha}$), our
observations constrain $\alpha$ to be greater than 1.0 in the $V$ band
(3$\sigma$ upper limit). This limit is consistent with the optical decay
found by \citet{Fox05b} using HST data between 5 and 10~d after the GRB
\citep[see also][]{Hjorth05b}. A similar limit was also put to the X-ray
afterglow decay. Flaring activity was reported in the X-ray light curve
\citep{Fox05b}. Given the limited available data, it is
difficult to say whether a similar behaviour was present also in the
optical band.
Our measurements on Jul 12 are nearly simultaneous with the first
\Chandra{} observation, so we can construct the broad-band spectral
energy distribution (Fig.~\ref{fig:sed}). The X-ray spectral index is
similar to that of long GRB afterglows \citep{DeP03,Nou05}. The
extrapolation of the X-ray spectrum matches the optical flux (that is,
the optical-to-X-ray slope $\beta_{\rm OX} = 1.1$ is consistent with the
X-ray slope $\beta_{\rm X} = 1.24 \pm 0.35$).
The spectrum corresponding to the optical colours is quite red
($\beta_{\rm opt} = 2.3 \pm 0.7$), but given its large uncertainty it is
consistent with $\beta_{\rm OX}$ at the 1.7$\sigma$ level. Small dust
extinction \citep[$A_V \approx 0.2$~mag;][]{Fox05b} would make the
intrinsic color bluer and fully consistent with the X-ray slope, so that
the optical and X-ray emission may constitute a single component.
However, the sparseness of the data prevents us from drawing any robust
conclusion.

\begin{figure}
\includegraphics[width=\columnwidth]{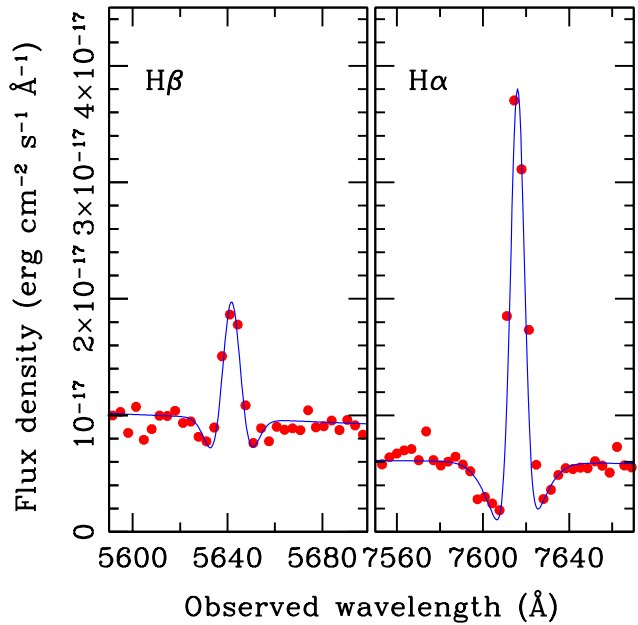} 
\caption{
Details of the GRB\,050709 host galaxy spectrum close to the H$\alpha$
and H$\beta$ lines, showing that the narrow emission emissions are
superimposed on wider absorption features.%
\label{fig:spe}}
\end{figure}

The properties of the host galaxy are intriguing. Its colours are
consistent with those of an irregular galaxy at $z \approx 0.2$
\citep{Fuk95}, and are much bluer than those of ellipticals (like those
associated with other short GRBs). Moreover, hints of morphological
structure are seen in our best-seeing images. A close inspection of the
spectrum shows that both the H$\alpha$ and H$\beta$ lines have a narrow
emission component (FWHM$\mbox{} < 6$~\AA) partially filling a wider
($\sim 12$~\AA{} FWHM) absorption feature (Fig.~\ref{fig:spe}). This
classical signature \citep{DG83} identifies a dominant stellar
population $\sim 1$~Gyr old (mostly A-dwarf stars), together with a
younger, hotter component. As indicated from the prominent nebular
emission lines, star formation is still present. From the H$\alpha$ and
[O\,II] emission lines (having luminosities $2.62 \times 10^{40}$ and
$2.45 \times 10^{40}$~erg~s$^{-1}$, respectively), we infer a star
formation rate of 0.21 and 0.34~$M_\odot$~yr$^{-1}$ \citep{Ken98}, which
corresponds to $\sim 2$--3.5~$M_\odot$~yr$^{-1}$ once normalised to
$L_*$. This is significantly less than that typically observed in long
GRB host galaxies \citep{Chr04}, but much larger than that in the hosts
of GRB\,050509B%
\footnote{However, many blue galaxies were located in the XRT
error circle.}
\citep{Blo05} and GRB\,050724 \citep{Beral05}, by factors of $> 50$ and
$> 150$, respectively.

The most popular model for short/hard GRBs is the merger of a binary
compact object system \citep[e.g.][]{Eic89}. Such events can occur in a
late-type, star-forming galaxy \citep{BeKa01}, and give rise to short
GRBs \citep{PeBe02}. Since the merging timescales may be of the order
of 10~Myr, small offsets between the explosion site and the galaxy core
are possible. Therefore, GRB\,050709 might have been produced in a
tightly bound system, with a short merging time, similar to GRB\,050724
\citep{Beral05}. However, we also note that according to the standard
Faber-Jackson relation, the escape velocity from the GRB\,050709 host is
quite large (about 300~km~s$^{-1}$), so that only a fraction of binary
systems may be able to escape its potential well. In this case, a larger
delay ($\sim 1$~Gyr) would be consistent both with the observed offset
and with the age of the older stellar population. A large instantaneous
star formation rate would not be expected in this case, even if this
does not pose any problem for the merger model.

\begin{figure}
\includegraphics[width=\columnwidth]{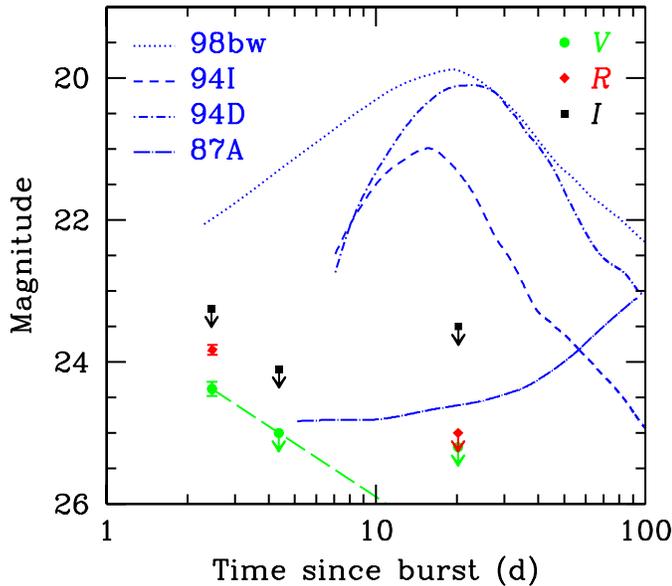} 
\caption{Light curve of the GRB\,050709 afterglow (points), compared
to those of several SNe ($R$ band). Zero extinction is assumed at the
GRB site.\label{fig:lc}}
\end{figure}

The presence of pronounced star formation activity in the host galaxy of
GRB\,050709, however, prompts us to investigate whether this event could
be directly related to young stars. Recently, it was proposed that short
GRBs may be produced by giant flares from soft gamma-ray repeaters
\citep[e.g.][]{Hurley05}. However, the luminosity of GRB\,050709 would
be a factor $\sim 10^3$ larger than that of the giant flare from
SGR\,1806-20. The prompt emission properties also make this hypothesis
unlikely \citep{Vill05}. Furthermore, our photometry can put strong
constraints on the presence of an unextinguished supernova (SN) exploded
simultaneously with the GRB (see Fig.~\ref{fig:lc}). Our limits impose a
SN $\ga 100$ times fainter than a typical type-Ia SN or a bright
hypernova like SN\,1998bw. Also fainter events like SN\,1994I and even
SN\,1987A are incompatible with our data. An association with a SN seems
therefore ruled out for GRB\,050709 \citep[see
also][]{Hjo05,Hjorth05b,Fox05b}. The properties of the GRB\,050709 host
are however consistent with the model proposed by \citet{MacFadyen05},
which advocates a collapsing neutron star accreting from a close
non-compact companion. Such model would also naturally explain the
flares observed in the X-ray light curve.

\begin{acknowledgements}
DM thanks INAF and the Italian MIUR for support. This research was
supported at OABr and OAR by ASI grant I/R/039/04. We acknowledge the
excellent support of the ESO staff. KH is grateful for support under
MIT-SC-R-293291 and FDNAG5-9210. We also thank the anonymous referee for
her/his valuable comments and suggestions.
\end{acknowledgements}

\end{document}